\documentclass[10pt,letterpaper]{article} 
\usepackage{opex3}
\begin{document}

\title{All-optical switching using Kerr effect in a silica toroid microcavity}
\author{Wataru Yoshiki and Takasumi Tanabe$^{*}$}
\address{
Department of Electronics and Electrical Engineering, Faculty of Science and Technology, Keio University, \\ 3-14-1, Hiyoshi, Kohoku-ku, Yokohama 223-8522, Japan\\
$^*$Corresponding author: takasumi@elec.keio.ac.jp
}

 \begin{abstract}
  We demonstrate experimentally an all-optical switching operation using the Kerr effect in a silica toroid microcavity. Thanks to the small mode volume and high quality factor of the silica toroid microcavity, we achieved on-chip optical Kerr switching with an input power of $2~\mathrm{mW}$. This value is the smallest among all previously reported on-chip optical Kerr switches. We also show that this value can be reduced to a few tens of $\mathrm{\mu W}$ by employing a mode with a $Q$ factor of $>2 \times 10^7$. 
 \end{abstract}

 \bibliographystyle{osajnl}{

 \section{Introduction}
 \noindent
 All-optical switches have been intensively studied with the expectation that they will dispense with opto-electro and electro-optic converters in telecom networks.
 All-optical switches based on optical microcavities are particularly attractive because they can be driven with an extremely low input power thanks to their high quality ($Q$) factor and small mode volume $V$.
 A high $Q/V$ is needed because these switches employ optical nonlinearity for their operation.
 Among various microcavity-based switches, the most prominent are those fabricated on a semiconductor substrate, such as photonic crystal (PhC) nanocavity switches \cite{Nozaki2010sfa,Tanabe2005fba,Husko2009uao} and microring switches \cite{Almeida2004aos,Waldow20082ao}, because of their high nonlinearity and their potential for large-scale integration on a chip \cite{Kuramochi2014lsi}.
 Microcavity-based switches often employ a carrier-induced effect \cite{Nozaki2010sfa,Tanabe2005fba} to modulate the refractive index.
 However, the response time of carrier-induced switching is limited by the carrier relaxation time.
 Moreover, these switches suffer from absorption loss originating from free carrier absorption (FCA).
 So, there is a need for a carrier-free microcavity-based switch.

 Recently, all-optical on-chip switches using the Kerr effect have been demonstrated by a number of groups \cite{Pelc2014pao,Eckhouse2012kia}.
 The Kerr effect is instantaneous and has a small loss.
 However, the threshold of the Kerr effect is usually higher than that of the carrier-induced effect \cite{Barclay2005nrs}; thus the demonstrated switches usually require a switching peak power of $>100~\mathrm{mW}$.
 To overcome this problem, optical Kerr switches were developed by using whispering gallery mode (WGM) microcavities.
 WGM microcavities typically have an ultra-high $Q$, thus they require an input power of less than $50~\mathrm{\mu W}$ \cite{Pollinger2010aos}.
 Switching has been demonstrated by using a hybrid silica microsphere \cite{Razdolskiy2011hmn}, a hydrogenated amorphous silicon (a-Si:H) microcylindrical cavity \cite{Vukovic2013uoc} and a silica bottle cavity \cite{Pollinger2010aos}.
 Although, they are not fabricated on a chip, they can be efficiently coupled to a tapered fiber \cite{Spillane2003iif,Knight1997pme}, and thus have a small coupling loss.
 This is a clear advantage for both optical telecommunication and loss-sensitive applications such as quantum information processing \cite{OShea2013fos,Aoki2006osc}.
 
 Of the various WGM microcavities, we are interested in the silica toroid microcavity \cite{Armani2003uhq}, because it is fabricated on a silicon chip and can be integrated with a waveguide \cite{Zhang2013smr}, which is not possible with other ultrahigh-$Q$ WGM cavities.
 Carrier generation is greatly suppressed in silica thanks to the large bandgap (corresponding to $\lambda = 140~\mathrm{nm}$) of the material, which enables us to use the Kerr effect.
 It has an ultra high $Q$ of $>10^8$ and a small mode volume of $<200~\mathrm{\mu m}^3$ \cite{Kippenberg2004duh}.
 The modulation of a signal pulse by the Kerr effect has been demonstrated by taking advantage of these properties \cite{Rokhsari2005okn}.
 Numerical analysis has shown that even a memory operation is possible \cite{Yoshiki2012abm}.

 In this paper, we demonstrate experimentally an all-optical switch using the Kerr effect in a silica toroid microcavity.
 The purpose of this study is to show how small an operating power a Kerr based switch can achieve.
 Indeed we will report successful operation at an input power of $2~\mathrm{mW}$, which is the smallest value for any previously reported on-chip optical Kerr switch \cite{Pelc2014pao,Eckhouse2012kia}.
 Moreover we will show that a minimum modulation power of $36~\mathrm{\mu W}$ is possible by employing a mode where $Q>2 \times 10^7$.
 
 This paper is organized as follows.
 In \S 2, we describe the fabrication of the microcavity and show the experimental setup we used for our all-optical switching demonstration.
 After that we show the results of the all-optical switching experiment and confirm that our optical switch is indeed based on the Kerr effect by comparison with the result of a numerical calculation.
 In \S3, we discuss the switching contrast and the required input power of our switch and compare the values with those of previously reported Kerr switches.
 Finally, in \S4, we summarize this study and conclude the paper.

 \section{Experimental results}\label{sec3} 
  \subsection{Device fabrication \& Experimental setup}
  We fabricated our silica toroid microcavity using (1) photolithography, (2) SiO$_2$ etching, (3) XeF$_2$ dry etching and (4) laser reflow \cite{Armani2003uhq}.
 The major and minor diameters were $70~\mathrm{\mu m}$ and $4.5~\mathrm{\mu m}$, respectively.
 A tapered fiber was used to couple the light into the microcavity \cite{Knight1997pme}.
 It was fabricated by heating and stretching a commercial single mode fiber so that it had a diameter of $\sim 1~\mathrm{\mu m}$.
 Our tapered fiber had a transmittance of about $90\mathrm{\%}$.
 
 Figure~\ref{fig:setup} shows a block diagram of our experimental setup.
 We used two laser lights, one as a control and the other as a signal.
 The wavelengths of the two laser lights $\lambda_\mathrm{in}^\mathrm{control}$, $\lambda_\mathrm{in}^\mathrm{signal}$ were tuned to the resonance of each mode $\lambda_\mathrm{res}^\mathrm{cont} = 1544.122~\mathrm{nm}$, $\lambda_\mathrm{res}^\mathrm{sig} = 1579.497~\mathrm{nm}$. 
 An EOM (10~GHz) was used to modulate the control, which had rectangular pulses.
 At the output, we used BPFs to filter out the control light.
 A PD was used to monitor the coupling efficiency of the control light into the microcavity.
 We performed all of the experiments under critical-coupling conditions so that we maximized the light energy in the microcavity.
 \begin{figure}[htbp]
  \begin{center}
   \includegraphics*[width=4in]{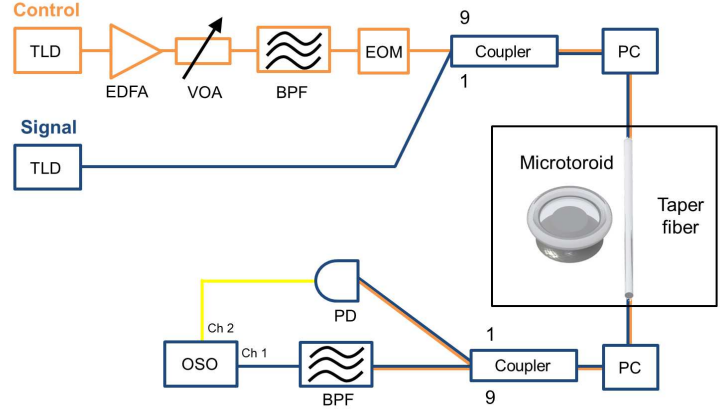}
   \caption{Experimental setup for all- optical switching.  TLD: Tunable laser diode, EDFA: Erbium doped fiber amplifier, VOA: Variable optical attenuator, BPF: Band pass filter, EOM:  Electro-optical modulator, PC: Polarization controller, PD: Photo diode, and OSO: Optical sampling oscilloscope.}
   \label{fig:setup}
  \end{center}
 \end{figure}
 
  \subsection{Demonstration of Kerr based all-optical switching}
  First, we describe our demonstration of all-optical switching.
  Although the material absorption is small in silica, the actual absorption of the silica-based microcavity is much larger \cite{Gorodetsky1996uqo,Rokhsari2004lci}.
  As a result, the TO effect is often larger than the Kerr effect when the system is in equilibrium, and it is necessary to distinguish the TO and Kerr effects.
  As shown in the references \cite{Rokhsari2005okn,Yoshiki2012abm}, the response time of the Kerr effect is around a few tens of $\mathrm{ns}$ (limited by the cavity photon lifetime), but that of the TO effect is much longer than $\mathrm{\mu s}$.
  So we can obtain Kerr switching when we input control pulses whose duration is much shorter than the response time of the TO effect.

  Figure~\ref{fig:switch_result} shows the results of all-optical switching with different control pulse durations.
  The loaded $Q$ factors of the control and signal modes ($Q_\mathrm{load}^\mathrm{cont}$,  $Q_\mathrm{load}^\mathrm{sig}$) are $3.3 \times 10^6$ and $5.1 \times 10^6$.
  When the modulation speed is slow, the signal recovery takes much longer than the cavity photon lifetime.
  Figure~\ref{fig:switch_result}(a) shows the modulated signal light when the input control peak power $P_\mathrm{in}^\mathrm{cont}$ is $1.9~\mathrm{mW}$ and the control pulse duration $T_\mathrm{cont}$ is set at $1~\mathrm{ms}$.
  The signal recovery time is $80~\mathrm{\mu s}$ (the time where the signal output is $1/e$ of its maximum), and it is clear from the speed that this switching is due to the TO effect.

  On the other hand, when we switch the cavity with $T_\mathrm{cont}=64~\mathrm{ns}$ and $P_\mathrm{in}^\mathrm{cont}=5.3~\mathrm{mW}$, we observe Fig.~\ref{fig:switch_result}(b).
  Now the signal recovery time is $6~\mathrm{ns}$, which is much shorter than the thermal response time of $>\mathrm{\mu s}$ \cite{Rokhsari2005okn}.
  Since the Kerr effect is instantaneous, the recovery time of the signal should be of about the same order as the sum of the photon lifetimes of the control and signal modes.
  To confirm this, we performed a numerical simulation based on coupled mode theory (CMT) \cite{Manolatou1999cma,Yoshiki2012abm} and calculated the behavior of the signal output after the control light was turned off.
  The result is shown as a red dotted line in Fig.~\ref{fig:switch_result}(b) and it agrees well with the experimental result.
  Therefore, we can conclude that we obtained all-optical switching thanks to the Kerr effect.
  
  As seen in Fig.~\ref{fig:switch_result}(b), the signal output temporarily exceeds ``1'' (corresponding to the off-resonance signal output).
  This is because the signal light stored in the cavity is immediately outputted in response to the resonance shift due to the control light.
  It should be emphasized that the average signal output never exceeds  ``1''.
  \begin{figure}[htbp]
   \begin{center}
    \includegraphics*[width=5in]{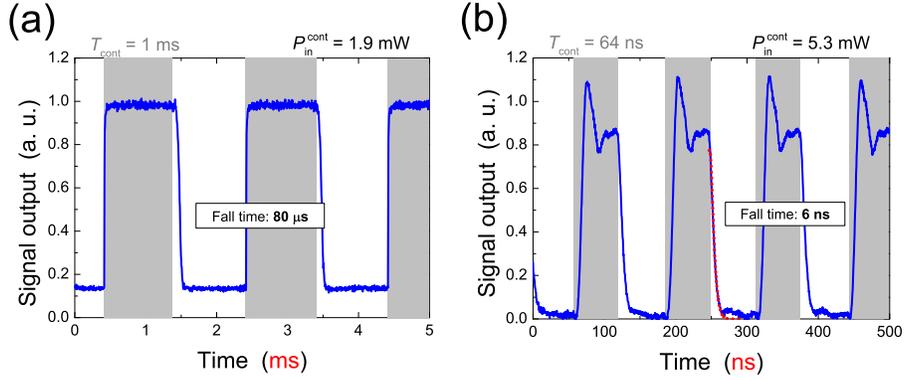}
    \caption{All-optical switching operation based on (a) TO and (b) Kerr effects. The solid blue line represents the signal output and the gray area indicates that the control light is inputted. The signal output is normalized by the off-resonance output. The red dotted line represents the signal output calculated by the simulation.}
    \label{fig:switch_result}
   \end{center}
  \end{figure}
  
 \section{Analysis and discussion}\label{sec} 
  \subsection{Influence of control pulse duration}
  Next, we examine the influence of $T_\mathrm{cont}$ in more detail.
  Since the response times for the TO and Kerr effects are different, there are two switching slopes in the signal output; the faster one is caused by Kerr effect and the slower one is caused by the TO effect.
  Figure~\ref{fig:switch_modulation}(a)-(d) show the switching operation for control pulses with different control pulse durations $T_\mathrm{cont}$.
   There is only a fast switching slope when the control pulse duration is shorter than $128~\mathrm{ns}$ (Figs.~\ref{fig:switch_modulation}(a) and (b)).
   Hence, the Kerr effect dominates the TO effect in this regime.
   However, we observe a slow decay when the pulse duration is longer than $512~\mathrm{ns}$ (Figs.~\ref{fig:switch_modulation}(c) and (d)).
   In this regime, we can observe both fast and slow switching slopes; thus both the Kerr and TO effects are responsible for the switching.
   This result tells us that we need to operate the system at a speed faster than $T_\mathrm{cont}$ at $<128~\mathrm{ns}$ if we are to use Kerr effect.
   \begin{figure}[htbp]
    \begin{center}
     \includegraphics*[width=5in]{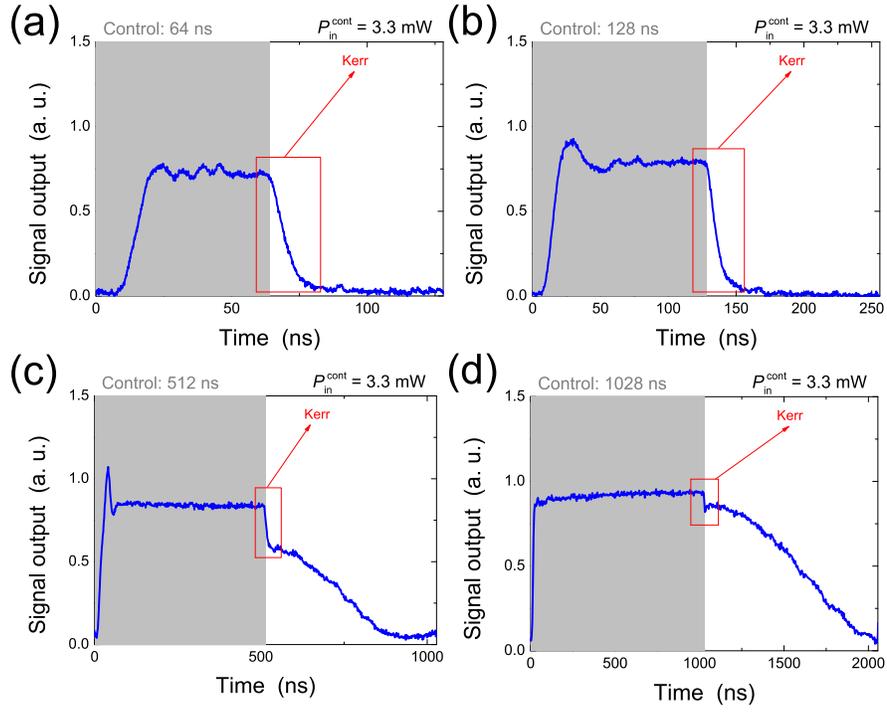}
     \caption{Signal output for control different pulse durations $T_\mathrm{cont}$. The blue line represents signal output when the signal wavelength is detuned to the resonance.}
     \label{fig:switch_modulation}
    \end{center}
   \end{figure}
   
   \subsection{Switching contrast versus switching power}
   Next we study the switching contrast.
   Figure~\ref{fig:switch_power}(a) shows the switching contrast for different input control powers.
   We calculate the switching contrast by dividing the modulated signal output by the off-resonance signal output (See the inset of Fig.~\ref{fig:switch_power}(a)).
   The blue dots are the experimental data, and the black curve shows the theory.
   The theoretical curve considers only the Kerr effect (it disregards TO effect) and is drawn by using Eqs.~(\ref{eq:nKerr_eq2}), (\ref{pump_probe_1}) and (\ref{eq:pump_probe_3}), which are shown in Appendix.
   In addition, we assume that both the control and the signal modes are fundamental WGMs (i.e. $(q,l) = (0,0)$).
   The spatial distributions of the WGMs are calculated by using the finite element method (COMSOL multiphysics) \cite{Oxborrow2007t2d}.
   Although we use parameters drawn from the literature and experiments (i.e. no fitting parameters), the experimental results agree well with the theory.
   This is further evidence that this switch operates with the Kerr effect.
   \begin{figure}[htbp]
    \begin{center}
     \includegraphics*[width=5in]{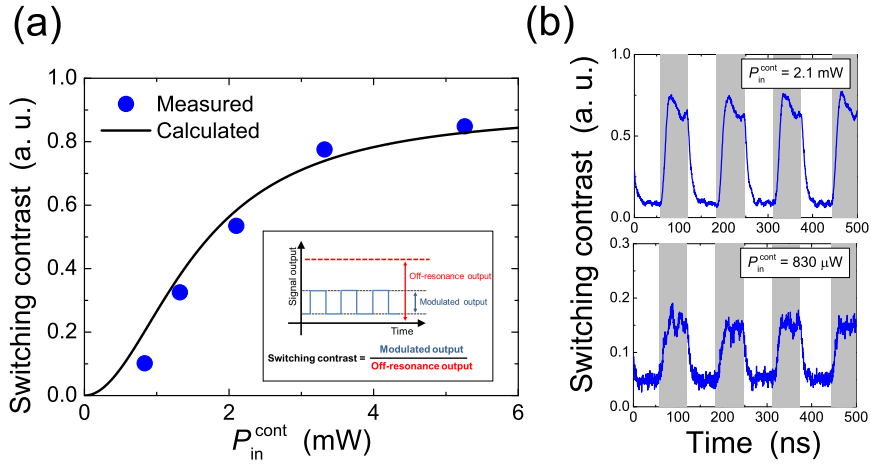}
     \caption{(a) Switching contrast $\eta$ versus input control power $P_\mathrm{in}^\mathrm{cont}$. The blue dots and black solid line are experimental data and a theoretical curve, respectively. The inset illustrates the definition of switching contrast. (b) The waveforms of signal outputs for different control powers ($P_\mathrm{in}^\mathrm{cont}=2.1~\mathrm{mW}$, $830~\mathrm{\mu W}$).}
     \label{fig:switch_power}
    \end{center}
   \end{figure}
   
   Figure.~\ref{fig:switch_power}(a) shows that an input control power of 5.3~mW is necessary to obtain a large switching contrast of 84\%.
   A contrast of about 3~dB is obtained at a power of 2.1~mW, and the modulation is still observed when we reduce the input control power to $830~\mathrm{\mu W}$.
   The waveforms for different control powers are shown in Fig.~\ref{fig:switch_power}(b), which reveals that the ON-OFF contrast is sufficient even for $830~\mathrm{\mu W}$.
  Since the low switching power is due to the high-$Q$ factor ($Q_\mathrm{load}>10^6$), we performed switching experiments using a silica toroidal microcavity with $Q_\mathrm{load}^\mathrm{cont}=4\times 10^7$ and $Q_\mathrm{load}^\mathrm{sig}=2\times 10^7$, as shown in Fig.~\ref{fig:switch_minimum}.
  Although the use of an ultra-high $Q$ factor makes the switching operation sensitive to various fluctuations, because of the narrower resonance of an ultra high-$Q$ cavity, we now observe modulation even at an input control power of $36~\mathrm{\mu W}$.
  \begin{figure}[htbp]
   \begin{center}
    \includegraphics*[width=3in]{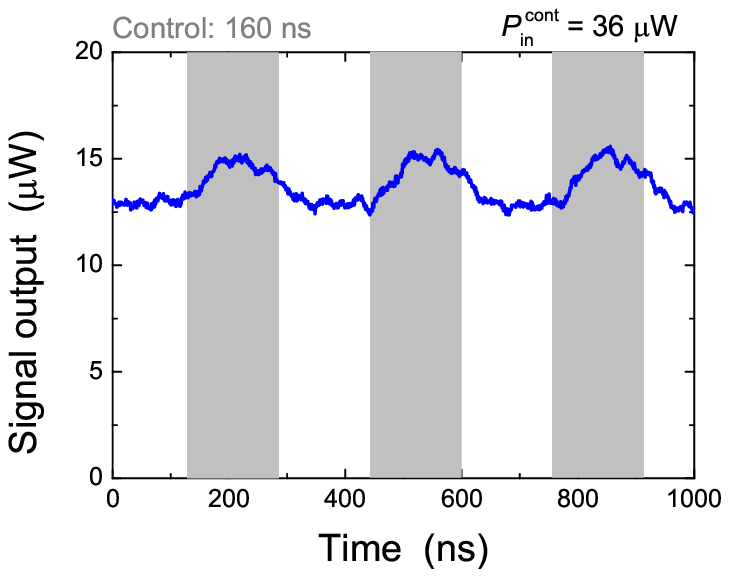}
    \caption{Signal output with an ultra high-$Q$ cavity. The signal output is amplified by an EDFA before being detected. The peak control power is $P_\mathrm{in}^\mathrm{cont} = 36~\mathrm{\mu W}$.}
    \label{fig:switch_minimum}
   \end{center}
  \end{figure}

  Finally, we compare the required control power of our switch with those of other optical Kerr switches.
  Table~\ref{tab:comp} summarizes the performance of various optical Kerr switches.
  It shows that the power of $36~\mathrm{\mu W}$ is the smallest of all the optical Kerr modulators \cite{Razdolskiy2011hmn,Pelc2014pao,Eckhouse2012kia,Vukovic2013uoc,Pollinger2010aos}.
  It should be noted this value is the raw power at the fiber.
  An optical switch fabricated on a semiconductor chip generally suffers from a large insertion loss, which is attributed to propagation loss and the coupling loss between a chip and an optical fiber.
  But our device has an extremely small loss, which is a key feature when using the switch for such applications as quantum information processing \cite{OShea2013fos,Aoki2006osc}.
  In addition, the silica toroid microcavity that we use for the platform of the optical Kerr switch can be fabricated on a chip \cite{Armani2003uhq} and it exhibits highly efficient coupling with a tapered fiber \cite{Spillane2003iif}.
  Therefore, our switch is advantageous in terms of on-chip fabrication and the negligible insertion loss in addition to the low required control power.
  \begin{table}[htbp]
   \caption{Comparison of the performance of optical Kerr switches.}
   \center
   \begin{tabular}{|l|l|l|l|}
    \hline
    Device & Peak control power in a fiber & Insertion loss \\ \hline
    Hybrid silica microsphere \cite{Razdolskiy2011hmn} & 1500~W & Small \\ \hline 
    a-Si:H microring \cite{Pelc2014pao} & 4.4~W$^{*)}$ & 12~dB \\ \hline 
    GaInP PhC Fabry-Perot \cite{Eckhouse2012kia,Tran2009pcm} & 800~mW$^{*)}$ & 13~dB \\ \hline 
    a-Si:H microcylinder \cite{Vukovic2013uoc} & 5~mW & Small \\ \hline 
    Silica microbottle \cite{Pollinger2010aos} & 50~$\mathrm{\mu W}$ & Small \\ \hline 
    This work & 36~$\mathrm{\mu W}$ & Small \\ \hline 
   \end{tabular} \\
   \footnotesize{$^{*)}$ These values are calculated taking the insertion losses of the devices into consideration.}
   \label{tab:comp}
  \end{table}
  
  \section{Conclusion}\label{sec4}
  In conclusion, we demonstrated an all-optical switching operation in a silica toroid microcavity using the Kerr effect.
  Thanks to the small mode volume and high $Q$ factor of the silica toroid microcavity, we developed an on-chip optical Kerr switch driven with an input control power of $2~\mathrm{mW}$.
  This value can be reduced to $36~\mathrm{\mu W}$ by using a higher-$Q$ factor and is the lowest of any of the previously reported optical Kerr modulators, which are based on, for example, an InGaP PhC cavity \cite{Eckhouse2012kia}, an a-Si:H microring \cite{Pelc2014pao}, and a silica microbottle \cite{Pollinger2010aos}.
  In addition, our switch is the only optical Kerr switch that can be fabricated on a chip and it exhibits highly efficient coupling with optical fiber.
  We believe that our low-power driven optical Kerr switch can be applied for both optical telecommunication and loss-sensitive applications such as quantum information processing.
  
 \section*{Acknowledgment} 
 This work is supported in part by the Strategic Information and Communications R\&D Promotion Programme (SCOPE) of the Ministry of Internal Affairs and Communications in Japan, the Support Center for Advanced Telecommunications Technology Research (SCAT), and the Leading Graduate School program for ``Science for Development of Super Mature Society'' from the Ministry of Education, Culture, Sports, Science, and Technology in Japan.
 
 \section*{Appendix A: Theory describing switching contrast}
 Here, we develop the equation that models the switching contrast of our optical switch.
 The Kerr effect is caused by the light energy stored in a microcavity.
 Thus, we first derive the relationship between the light energy in the cavity $U_\mathrm{cavity}$ and the input control power $P_\mathrm{in}^\mathrm{cont}$ by using CMT \cite{Manolatou1999cma,Yoshiki2012abm}, as,
 \begin{eqnarray}
  U_\mathrm{cavity} = \left( 1 \mp \sqrt{Tr_\mathrm{res}^\mathrm{cont}} \right)\frac{\lambda_\mathrm{res}^\mathrm{cont} Q_\mathrm{load}^\mathrm{cont}}{\pi c}P_\mathrm{in}^\mathrm{cont}, \label{eq:Ucavity_Pin} 
 \end{eqnarray}
 where $c$, $\lambda_\mathrm{res}^\mathrm{cont}$ and $Tr_\mathrm{res}^\mathrm{cont}$ are the light velocity, the resonant wavelength and the transmittance on resonance of the control mode, respectively.
 $Q_\mathrm{load}^\mathrm{cont} = (1/Q_\mathrm{int}^\mathrm{cont} + 1/Q_\mathrm{coup}^\mathrm{cont})^{-1}$ is the loaded $Q$ factor of the control mode, where $Q_\mathrm{int}^\mathrm{cont}$ and $Q_\mathrm{coup}^\mathrm{cont}$ are the intrinsic $Q$ factor and the $Q$ factor related to the coupling between the waveguide and the cavity, respectively. 
 Note that the upper and the lower signs of Eq.~(\ref{eq:Ucavity_Pin}) correspond to the under- and over-coupling conditions, respectively.
 We assumed that the wavelength of the control light $\lambda_\mathrm{in}^\mathrm{cont}$ is tuned to the resonant wavelength of the control mode $\lambda_\mathrm{res}^\mathrm{cont}$ and we neglected the light energy stored due to the signal light.
 
 Next, we obtain an equation that will gives the relationship between the resonant wavelength shift of the control mode $\delta \lambda^\mathrm{cont}$ and the input control power $P_\mathrm{in}^\mathrm{cont}$.
 According to Ref.~\cite{Yoshiki2012abm}, the refractive index change caused by the Kerr effect $\Delta n_\mathrm{Kerr}(r,z)$ is written as,
 \begin{eqnarray}
  \Delta n_\mathrm{Kerr}(r,z) = \frac{2 n_2 c}{n_0}u_\mathrm{cavity}(r,z) = \frac{n_2 c}{n_0 \pi R} U_\mathrm{cavity} \tilde{I}(r,z),
   \label{eq:nKerr_eq1}
 \end{eqnarray}
 where $n_2$, $R$, $n_0$, $\tilde{I}(r,z)$ and $u_\mathrm{cavity}(r,z) = U_\mathrm{cavity}/(2\pi R) \cdot \tilde{I}(r,z)$ are the nonlinear refractive index, the major radius of the cavity, the refractive index, the normalized WGM intensity ($\int \int \tilde{I}(r,z)\mathrm{d}r\mathrm{d}z=1$) and the light energy distribution in the cross-section of the cavity, respectively.
 By using Eq.~(\ref{eq:nKerr_eq1}), we obtain the wavelength shift of the control mode induced by the Kerr effect $\delta \lambda^\mathrm{cont}$ as, 
 \begin{eqnarray}
  \delta \lambda^\mathrm{cont} = \frac{\lambda_\mathrm{res}^\mathrm{cont}}{n_0} \int \int \Delta n_\mathrm{Kerr}(r,z) \tilde{I}(r,z)\mathrm{d}r\mathrm{d}z. \label{eq:dl_Kerr}
 \end{eqnarray}
 This equation considers the spatial overlap between the distributions of the refractive index change and the WGM \cite{Kippenberg2004kno}.
 Then, we derive the complete equation for $\delta \lambda^\mathrm{cont}$ by combining Eqs.~(\ref{eq:Ucavity_Pin}), (\ref{eq:nKerr_eq1}) and (\ref{eq:dl_Kerr}), as,
 \begin{eqnarray}
  \delta \lambda^\mathrm{cont} = \frac{(\lambda_\mathrm{res}^\mathrm{cont})^2 n_2 Q_\mathrm{load}^\mathrm{cont}}{n_0^2 \pi^2 R}\left( 1 \mp \sqrt{Tr_\mathrm{res}^\mathrm{cont}} \right) P_\mathrm{in}^\mathrm{cont} \int \int \tilde{I}^2(r,z)\mathrm{d}S. \label{eq:nKerr_eq2}
 \end{eqnarray}
 
 Here, we obtain the change in the transmittance of the signal light induced by the control light.
 The wavelength shift of the control mode reflects that of the signal mode, thus the wavelength shift of the signal mode $\delta \lambda_\mathrm{cont}$ is written, as,
 \begin{eqnarray}
  \delta \lambda^\mathrm{sig} = 2\delta \lambda^\mathrm{cont}. \label{pump_probe_1}
 \end{eqnarray} 
 We multiplied $\delta \lambda^\mathrm{cont}$ by ``2'' because this is the cross-phase modulation (XPM) \cite{Agrawal1995nfo}.
 By using~CMT \cite{Manolatou1999cma,Yoshiki2012abm}, the transmittance of the signal light $Tr^\mathrm{sig}$ is written as,
 \begin{eqnarray}
  && Tr^\mathrm{sig} = \frac{\Delta^2 +  \left\{ \frac{\pi c}{\lambda_\mathrm{res}^\mathrm{sig}}\left(\frac{1}{Q_\mathrm{int}^\mathrm{sig}} - \frac{1}{Q_\mathrm{coup}^\mathrm{sig}} \right) \right\}^2}{\Delta^2 + \left( \frac{\pi c}{\lambda_\mathrm{res}^\mathrm{sig} Q_\mathrm{load}^\mathrm{sig}} \right)^2} = \frac{\Delta^2 +  \left( \frac{\pi c}{\lambda_\mathrm{res}^\mathrm{sig} Q_\mathrm{load}^\mathrm{sig}} \right)^2 Tr_\mathrm{res}^\mathrm{sig}}{\Delta^2 + \left( \frac{\pi c}{\lambda_\mathrm{res}^\mathrm{sig} Q_\mathrm{load}^\mathrm{sig}} \right)^2},\label{eq:pump_probe_22} \\
  && Tr_\mathrm{res}^\mathrm{sig} = Tr^\mathrm{sig}(\Delta = 0) = \left\{Q_\mathrm{load}^\mathrm{sig} \left( \frac{1}{Q_\mathrm{int}^\mathrm{sig}} - \frac{1}{Q_\mathrm{coup}^\mathrm{sig}} \right)\right\}^2, \label{eq:pump_probe_23} \\
  && \Delta = 2\pi c \left(\frac{1}{\lambda_\mathrm{res}^\mathrm{sig} + \delta \lambda^\mathrm{sig}} - \frac{1}{\lambda_\mathrm{in}^\mathrm{sig}}\right), \label{eq:pump_probe24}
 \end{eqnarray}
 where $T_\mathrm{res}^\mathrm{sig}$ is the transmittance of the signal mode when the signal wavelength is tuned to the resonance ($\lambda_\mathrm{in}^\mathrm{sig}=\lambda_\mathrm{res}^\mathrm{sig}$) and there is no resonance shift ($\delta \lambda^\mathrm{sig}=0$).
 Equation~(\ref{eq:pump_probe_22}) shows that if the wavelength of the signal mode is shifted ($\delta \lambda_\mathrm{sig} > 0$), the transmittance $Tr^\mathrm{sig}$ is also changed.
 When we initially set the signal wavelength $\lambda^\mathrm{sig}_\mathrm{in}$ at the resonance $\lambda_\mathrm{res}^\mathrm{sig}$, the switching contrast $\eta$ is defined as follows:
 \begin{eqnarray}
  \eta = \frac{\Delta^2 +  \left( \frac{\pi c}{\lambda_\mathrm{res}^\mathrm{sig} Q_\mathrm{load}^\mathrm{sig}} \right)^2 Tr_\mathrm{res}^\mathrm{sig}}{\Delta^2 + \left( \frac{\pi c}{\lambda_\mathrm{res}^\mathrm{sig} Q_\mathrm{load}^\mathrm{sig}} \right)^2} - Tr_\mathrm{res}^\mathrm{sig}. \label{eq:pump_probe_3} 
 \end{eqnarray}
 By using Eqs.~(\ref{eq:nKerr_eq2})-(\ref{eq:pump_probe_3}), we obtain the switching contrast.
\end{document}